\documentclass[preprint]{jpsj2}
\title{Flexible Foraging of Ants under Unsteadily Varying Environment}

\author{\textsc{Tomomi Tao}$^{1}$\thanks{E-mail address: tao@ms.osakafu-u.ac.jp},{
Hiroyuki Nakagawa}$^{1}$ \textsc{Masato Yamasaki}$^{2}$ and
\textsc{Hiraku Nishimori}$^{1}$\thanks{E-mail address: nishimor@ms.osakafu-u.ac.jp}}

\inst{$^{1}$Department of Mathematical Sciences, Osaka Prefecture University,Sakai 599-8531. \\
$^{2}$Department of Mathematical Sciences, Ibaraki University,Mito
310-8512. \\
}

\abst{
Using a simple model for the trail formation of ants,  
the relation between  i)the schedule of feeding which 
represents the unsteady natural environment, ii)emerging patterns of trails
connecting a nest with food resources, and iii)the foraging efficiency 
is studied.  Simulations and a simple analysis show that the emergent trail pattern flexibly 
varies depending on the feeding schedule by which 
ants can make an efficient foraging according to the 
underlying unsteady environment.
}

\kword{pattern formation, self-organization, ants}

\begin{document}
\maketitle

\section{Introduction} 

So far, various studies have been made on the collective behavior
of social insects \cite{Wil,Got,Hae,De1,De2,Hel,Schw1,Schw2,Bo,Sug1,Sug2,Nishinari,Chow}. 
Especially, the trail formation of ants \cite{Wil,Got,Hae,De1,De2,Hel,Schw1,Schw2,Bo,Nishinari,Chow} 
have called a wide interest as a remarkably synergetic 
behavior by
presumably unconscious individuals \cite{Hak,Pri}.
Still, many aspects are left unconfirmed on the 
trail formation process, like the detailed ingredient of almost all
types of chemicals(pheromone) \cite{Wil,Got} secreted by individual ants and their 
roles in each sociobiological regime \cite{Wil,Got,Hae,De1,De2}. 
In spite of such inadequate understanding of the reality,
various models on ants \cite{De2,Hel,Schw1,Schw2,Sug1,Sug2,Nishinari,Chow} 
have been proposed to imitate their behavior. 
Here,  setting a simple set of rules of which several ingredients 
are following previous studies \cite{De2,Hel,Schw1,Schw2,Sug1,Sug2,Nishinari,Chow}, we 
investigate the relation between i)the schedules of feeding which 
represent the unsteady natural environment, ii)emerging patterns of trails
which connect between a nest with food resources, and iii)the foraging efficiency 
as a group. Our final purpose is to connect the pattern formation of the trail
to the group strategy of foraging ants under an unsteadily varying environment.

In the below, the outline of the present model is explained. 
As mentioned above, several  parts of the present dynamics follow previous studies\cite{De2,Hel,Schw1,Schw2} 
at least in a qualitative sense, and it should be stressed that the main purpose of this paper is not to introduce
an essentially novel model of swarming ants,  
but by controlling the feeding schedule using a proper set of control parameters, 
we connect the emerging patterns of trails to the foraging efficiency of ants and 
discuss on the flexible tactics of them according to the unsteadily varying environment.

%As mentioned above, several  simulations with similar dynamical rules to the present ones 
%have already been introduced to imitate the  
%foraging behavior of ants, and some parts of them have reproduced the formation of 
%branching trails though not so systematically studied as the present simulation with the relation  to the feeding schedule.
%In that sense, the dynamical rules of ants in the present models are qualitatively  
%difference  from that in the preceding studies except for some details.  
%In that situation,  the main purpose of this study is not to introduce qualitatively different dynamics of ants from the preceding 
%models,  but to connect the foraging efficiency of ants to the pattern dynamics of trails which flexibly changes 
%according to the unsteady feeding environment.
 
\section{MODEL}

The Model consists of a colony of N movable agents (we call them 'ants') 
which are situated on a 2D honeycomb lattice with periodic boundary
condition. Each of these ants is located at one site in the lattice   
and is heading to one of the six nearest sites(fig. 1) to which, the right-hand neighbor or the left-hand, she will move according to the 
dynamical rule described below.  
The only task for ants is foraging, that is, finding food and carrying
it back to a nest and the feeding sites are located at two corners of a
regular triangle with edge-length $L$ and the nest as the third corner
(fig. 2).
The amount and the schedule of feeding are 
controlled through a pair of control parameters.
In particular, at every constant interval $T$, 
a certain amount $M$ of food is supplied at either of two feeding sites 
by turn(fig. 2).

Until the next feeding,
the amount of food at the feeding sites  
monotonically decreases as ants carry them away.
Notice that the upper limit of the temporal amount of food $M(t)$ at the feeding sites is set as $M$, 
that means if the residual amount of food just before the feeding time is not zero the additional food is diminished so that the total amount becomes $M$ after the addition. 
As the mean for the communication among ants, two types of attractive 
chemicals, pheromones, are secreted and perceived by individual 
ants according to their temporal modes which consist of: i)the semi-random walk mode (mode-I), ii)the exploring mode (mode-II) and 
iii)the homing mode (mode-III).
The kind of pheromones considered in the present model consists of: 
i)recruit pheromone and ii)foot pheromone,
with densities  $\rho_{\textup rec}({\bf x},n)$ and
$\rho_{\textup foot}({\bf x},n)$, respectively, where {\bf x} is the position of the 
underlying site and $n$ is the corresponding Monte Calro time step \cite{Tao}. 
The detailed behaviors of ants in individual modes
and the transition rule between different modes are the following:

\noindent
[mode-I: semi-random walk mode]Mode-I is a semi-random walk mode. In this mode
every ant starts/restarts from the nest, thereafter remains in this mode until she reaches a site with more than a critical density of recruit pheromone such that $\rho_{\textup rec}({\bf x},n) > \rho_{c} $.
More specifically, each ant is, in this mode, at each time step, 
heading in one of the six nearest sites(fig. 1).
This heading direction corresponds to the moving direction in the last
walking step while the moving direction in the next step is
the forward (=heading) direction or one of its neighboring (the right and the left)
directions. To choose one of them a stochastic rule with an equal weights
is applied. Note that only three neighboring sites are permitted as the next position 
among six nearest neighbors, therefore, we call the mode-I a semi-random walk mode.  
In this mode after each walking event, 
a time-varying amount of foot pheromone $\rho_{\textup foot}^{\textup add}(n')$ is secreted at the corresponding site, 
where $\rho_{\textup foot}^{\textup add}(n')$ decreases as $\rho_{\textup foot}^{\textup add}(n'+1)=B_{\textup foot}\rho_{\textup foot}^{\textup add}(n')$ 
depending on $n'$; the number of moving steps the ant has experienced after it started/restarted from the nest, 
where a decreasing rate $0<B_{\textup foot}<1$ is assumed. 
With such behavior by a crowd of ants, a roughly monotonic gradient field of foot pheromone 
is formed around the nest which 
serves as the landmark for homing(mode-III) ants though some degrees of its  deformation
is unavoidable because of the behavior of ants in mode-II who also secrete
the same kind of pheromone as explained soon.
On reaching the site
with $\rho_{\textup rec}({\bf x},n) > \rho_{c} $, its mode is 
excited into mode-II, or, on arriving at sites with food, 
her mode changes into mode-III

\noindent
[mode-II: exploring mode]Mode-II is the exploring mode for food where 
ants walk perceiving the local gradient of 
recruit pheromone density which is secreted by mode-III ants to 
call up other ants to the food resources.
The moving rule of ants in this mode is
the same as that in the mode-I except for the weight
$P_\alpha = exp(-\Delta^\alpha/T)/Z$ in the stochastic rule  
where $\Delta^\alpha \equiv \rho^\alpha-\rho({\bf x},t)$ 
is the gradient of pheromone density to individual directions,
$Z$ is the normalization factor and $\alpha$ indicates 
forward, right, or left direction  
relative to the heading  direction. 
In this mode, like in the mode-I, after each walking step, 
a time-varying amount of foot pheromone $\rho_{\textup foot}^{\textup add}(n')$ is secreted at the corresponding site.
On arriving sites with food, 
the mode changes into mode-III, or, 
once an ant gets lost from the activated sites with $\rho_{\textup rec}({\bf x},n) > \rho_{\textup c}$,
her mode turns back to the previous mode-I.

\noindent
[mode-III: homing mode]Mode-III is the homing mode after obtaining food.
On arriving at food, regardless of her previous mode,
an ant switches her mode into III in which, perceiving the local 
gradient of foot pheromone secreted by the mode-I and the mode-II ants, she walks
back to the nest carrying a unit amount, $m=1$, of food.  During this mode the recruit 
pheromone is secreted at each site in the walk with the amount decreasing according to the equation
 $\rho_{\textup rec}^{\textup add}(n'+1)=B_{\textup rec}\rho_{\textup rec}^{\textup add}(n')$ with $0<B_{\textup rec}<1$, here
$n'$ indicates the number of walking steps of the ant after 
leaving the feeding site.
On arriving at the nest her mode is switched back to the mode-I to restart for foraging. 

In addition to being secreted by ants, pheromones evaporate/decay at site {\bf x} according to
\begin{equation} 
\rho_\beta({\bf x},n+1)-\rho_\beta({\bf x},n)=-A_\beta\rho_\beta({\bf x},n)
\end{equation}
where $A_\beta$ are the evaporation rates and  $\beta$ is the index of 
recruit/foot pheromones.
Moreover, pheromones diffuse to the nearest sites\cite{Diff} 
\begin{equation} 
\rho_\beta({\bf x},n+1)-\rho_\beta({\bf x},n)=D(\langle \langle \rho_\beta({\bf x},n) \rangle \rangle -\rho_\beta({\bf x},n))
\end{equation}
where $\langle \langle \rangle \rangle$ means the average over the nearest sites of 
${\bf x}$, and $D$ is a diffusion constant.
Furthermore, ants get 'energy' on reaching the food site or 
the nest. The value $F_{\beta}$ of this energy decrease as they walk according to
$F_\beta(n+1)=B'_{\beta}F_{\beta}(n)$ where $0 < B'_{\beta} < 1$.
Here the decreasing rate is the same as that of the amount of the corresponding pheromone secretion, 
i.e., $B'_{\beta}=B_{\beta}$ where  $\beta$ is the index of 
recruit/foot pheromones.
If the energy falls to zero the ant is forced to go back to the nest to restart the
foraging behavior in the mode-I. 
This rule is introduced to avoid the endless semi-random or 
circular walks which are hardly observed in nature\cite{Wil}.

\section{SIMULATION}
All the simulations are performed on a $150 \times 150$  honeycomb lattice 
with periodic boundary condition. 
At the initial time step, N=500 ants are located at the nest, 
and no pheromone is distributed in the field when all ants are 
simultaneously released from the nest. Note the unit time 
in our simulation corresponds to one Monte Calro step within which 
N randomly chosen ants will sequentially take one individual step. 
Under a proper combination of fixed parameters mainly related to the behavior of ants
as summarized in Table I,
with a pair of control parameters $\{M,T\}$ 
related to the schedule of feeding, the formation process of trail patterns and 
the subsequent efficient foraging are investigated.
We are particularly interested in the relation between i)the geometries of trails 
and ii)the accompanying foraging efficiency.

Before showing the several characteristic 
geometries of trails and the conditions for 
the appearance of each of them,
we introduce (or reconfirm) three different time scales particularly
relevant to this model: 

\noindent
i)$T$: the feeding interval, i.e., the interval between two successive feeding events
which take place at alternate feeding sites. 

\noindent
ii)$T'$: the residence time of food. 
This is the period after a major part of the ants begin carrying food at one feeding site 
along a trail until all the food is taken away.
%assumig a straight trail is extending between the nest and a feeding site. 
Note $T'$ itself is not a directly controlled quantity, instead, it is 
defined through $M$ as $T'\equiv a^{-1}M$ where $a$ is the potentially maximum foraging efficiency $E_{\textup max}$, 
which is obtained under an ideal condition such that food is permanently supplied at one of two feeding sites 
and a straight trail between the feeding site and the nest is kept. The specific form of $a$ is $a=\gamma \frac{Nmv}{2L}$ where $m=1$ is the load capacity of food, $v=1$ is its walking velocity, $L=30$ is the distance between the nest and an feeding site along a straight path, $N=500$ is the total number of ants, and 
$\gamma$ is the net ratio of foraging ants 
who are shuttling between the nest and food which is inevitably smaller than 1 because of the substantial randomness of the Monte Carlo method.
The parameter  $\gamma$ is roughly estimated as 0.72 by the simulation explained later. 
Note that food is not completely carried away before the next feeding if $M$ is sufficiently large.
In this case actual residence time goes to infinity, however we still define 
the residence time as $T'=a^{-1}M$ which has a finite value.
Note a similar quantity $T'_{\textup Y}$ will be defined when a foraging procedure along the Y-shaped trail is considered 
where the quantity $a$ is replaced by $a'=\gamma \frac{Nmv}{2L'}$', here $L'(>L)$ is the path distance
between the nest and a feeding site along one branch of the Y-shaped trail.  

\noindent
iii)$\tau$: the trail formation time, i.e., a characteristic time for ants to construct a 
new trail from the nest to a feeding site. Remark that in this period the exploring 
time for new food and the calling up time of the sufficient number of ants to establish a stable trail are included.
The specific form of $\tau$ is unknown but should be a function of several parameters introduced above.
In the present study, of the particularly importance is $\tau$'s dependency on $L$; the path distance between the nest and the food along a trail, 
and $\tau$ should be an increasing function of $L$ considering 
its definition and the dynamics of the modeled ants.
Note that a similar quantity $\tau_{\textup Y}(<\tau)$ will be introduced which is the formation time for a branch of a Y-shaped trail.

In addition to the above time scales, the residence time of two type of pheromones
are of the great importance. Certainly they are subject to evaporation and diffusion rates. 
However, the relevant effect in the present simulation are the residence time of the accumulated 
pheromones secreted by a mass of ants in the foraging process. 
In the present simulation these are roughly estimated to be equal or longer than $T$ for the foot pheromone and much shorter than $T$ for the recruit pheromone. Note they vary strongly depending on the regime of foraging.

\subsection{A. Trail geometry}
Once a certain shape of trail is established, irrespective of temporal fluctuation, it is roughly classified into three shapes: i)V-shaped trail, ii)Y-shaped trail
or iii)/-shaped trail as shown in fig. 3.
Notice that these shapes are recognizable and definable after averaging the density field of ants over a period much longer period 
than the feeding period $T$.  The detailed conditions for the appearance of each of these trails are like followings:

\noindent 
i)V-shaped trails: When the amount $M$ of food supplied at each feeding event is inadequate to remain until the next feeding, while the residence time $T'\equiv a^{-1}M$  is longer than the trail formation time $\tau$, a straight trail connecting the nest and the last feeding sites 
is temporarily established and, thereafter, dissolves as food resource is exhausted at the site.
After new food is supplied at the alternative feeding site, a straight trail to
the new site is built to be maintained until the food resource is exhausted.
The same process is repeated after each feeding event.
Remember the feeding site switches from one to the other by turns so that when 
the trails are averaged over sufficiently longer than $T$, 
they can be recognized as the V-shaped trail.
In terms of the three characteristic time scales,
this type of trails is typically seen when  $T >> T'(=aM) > \tau $ within 
the parameter range of the present simulation, 
where the latter inequality is required to guarantee the formation of 
trails.

\noindent
ii)/-shaped trails: Typically when $T'(=aM) > 2T$ holds, namely, the amount of supplied food at each feeding event is 
adequate enough to remain until the next feeding, a  
single straight trail connecting the nest and one feeding site is 
kept. We classify this type of trail as the single straight trail
and describe it as the /-shaped trail. 
This type of trail also emerges at $T'(=aM) < 2T$ if the difference between $T'$ and $2T$
is not so large compared to $\tau$.

\noindent
iii)Y-shaped trails: At the intermediate region between i) and ii)
the Y-shaped trail is formed.
Here the amount of supplied food at respective feeding sites are 
inadequate to offer a permanent food source, however unlike the cases with the V-shaped trails, 
the amount is not so little as to completely dissolve the trail before a new trail to the alternative feeding site is built for the new food resource. 
In this situation, ants, in the beginning, build a V-shaped trail like in the above case, but as time proceeds it will 
deform into a Y-shaped trail which has a junction (i.e., a branching point at the midway) used as a 
'base-camp' to explore alternative food resources after one feeding site is exhausted,
while the {\it trunk trail} extending from the nest to the junction is kept\cite{trail}.
As explained later, the formation of the Y-shaped trail is remarkable in view of the efficient foraging.  
Note that the position of the junction in the Y-shaped trails continuously varies according to the control parameters $\{M,T\}$ and the V-shape is taken as an extreme case of the Y-shape. Therefore,
a strict distinction between these two shapes (the Y-shape and the V-shape) of trails is difficult.  
However, in practice, the {\it transient}  Y-shape with a very short vertical part ({\it the trunk trail})
is seen only in a narrow range of control parameters, and in the dominant cases the emerging Y-shape has
a junction situated almost equidistant from the two feeding sites and the nest. 
Thus, it is possible to roughly distinguish between these shapes 
according to the typical conditions within they emerge.
 
The obtained features of trails in the simulations and their dependence on the 
control parameters $\{M,T \}$ are displayed in fig. 4 
which depicts the spatial distribution of ants averaged over a time span significantly longer than $T$.

As can be seen from this figure, the most relevant quantity for the trail geometry seems to be  $M/T({\textup i.e.,} aT'/T)$. 
As $M/T$ increases, the shape of trail 
changes roughly from the V-shape to the Y-shape and finally the /-shaped
trail is obtained. 
In addition, $T$ and $T'$ are individually relevant for the 
geometry of trails. 
For example, in fig. 4, if one keeps $M/T$ constant (e.g. at 5/2), with the increase of $T$ the position of the junction of the 
Y-shaped trail in each figure gradually goes down to the nest,
finally is degenerates into the V-shape.

\subsection{B. Foraging efficiency}
To investigate the relationship between the efficiency of foraging 
and trail geometry, the averaged foraging efficiency $E_{\textup av}$ is introduced 
which is the averaged amount of food
carried into the nest per unit time.
Then, the relation between $M/T$ and  $E_{\textup av}$ 
is measured under various feeding rate $M/T$ with $T$ fixed. Fig. 5 shows the case for $T=600$. In this figure,
with a certain degree of fluctuations, a characteristic relation between $M/T$ and $E_{\textup av}$ is seen 
until $E_{\textup av}$ reaches an saturation value.

In this $(\frac{M}{T},E_{\textup av})$ relation, 
considering the emergent geometry of trails, roughly five characteristic zones 
are recognized as indicated by symbols $A, A', B, C$ and $D$.
In zone $A$ with the emergence of the V-shaped tail, an almost linear $(\frac{M}{T},E_{\textup av})$ relation is seen,   
and its slope is 1 which corresponds to the {\it perfect foraging} 
where ants collect all the supplied food at both feeding sites 
along the V-shaped trail.
On the other side, in zone $D$, $(\frac{M}{T},E_{\textup av})$ relation reaches a plateau, where 
the averaged foraging efficiency $E_{\textup av}$ amounts to the potentially maximum foraging efficiency $E_{\textup max}\equiv \gamma \frac{Nmv}{L} \equiv a$ which was introduced in the beginning of this section. This situation is realized if the amount of supplied food at each feeding site exceeds or is equal to the carrying capacity of ants $M/2T \ge E_{\textup max}$ where the equivalence holds at the left edge of zone $D$.
These two zones $A$ and $D$ are characterized by simple foraging strategies in such a sense that
ants make a straight trail (for the /-shaped trail) or a pair of straight trails(for the V-shaped trail) 
which realizes the most efficient foraging in time period
shorter than $T'$ and $T$. 
Conveniently, the continuous/alternate use of this/these straight trail/trails 
realizes the long-term optimized foraging over sufficiently longer periods than $T$.
Therefore, no conflict emerges between the short-term foraging strategy and 
the long-time foraging strategy. 

On the other hand, out of these two zones, a certain degrees of conflict arises between the 
short-term strategy and the long-term strategy:
For example, as $M$ increases over a critical value $M_{\textup C1}$ in zone $A'$, 
if ants stick to the previous strategy with the V-shaped trail, the sum 
$T'+\tau$ exceeds $T$. It means that a {\it basic period} $T$ between two successive feeding events finishes  
before the supplied food at the most recently fed site is completely carried away. 
Because $T'\equiv M/a$, the critical value for this {\it imperfect foraging} by the V-shaped trail is estimated as $M_{\textup C1} =a(T-\tau)$. 
However, what is seen in the simulation at zone $A'$ is, instead of sticking to the V-shaped trail formation, 
the formation of a Y-shaped trail which serves the system with a shorter formation time 
$\tau_{\textup Y}$ than $\tau$ of a new path to the alternative feeding site. 
This is to develop(rebuild) a new(previously formed) branch starting not from the nest but from the middle of the last trail while 
leaving the {\it trunk trail} (the vertical part of Y) steadily extending between the nest and the junction.
Here $\tau_{\textup Y}$ is defined as the formation time of a new branch (an oblique part of Y)  
which extends between the junction and either of two feeding sites.
 As a result, the {\it perfect foraging} is kept realized over $M=M_{\textup C1}$ up to $M=M_{\textup C2}$, the right edge of zone $A'$.
Notice that the Y-shaped trail, while saving the trail formation time to the new food resources,
causes a temporarily redundant foraging in such sense that the potentially maximum foraging efficiency for 
the Y-shaped trail does not reach to
$E_{\textup max}$, but falls to $ E_{\textup max}(Y) \equiv \gamma \frac{Tmv}{L'}  \equiv a_{\textup Y} (<a)$ because of 
the longer path distance $L'(>L)$ between the nest and a feeding site along the Y-shaped trail than that along the V-shaped trail.
Therefore, the alternation of the trail geometry between V and Y  
is taken as a {\it trade-off strategy} between that of saving the trail formation time to 
the alternative food source, and that of minimizing the path distance from the nest to food sources.
With simple arguments explained below this trade-off is advantageous for the Y-shaped trail 
if $M > M_{\textup C1}=(T-\tau)a$ as long as 
$T<\frac{\tau a-\tau_{\textup Y} a_{\textup Y}}{a-a_{\textup Y}} \equiv T_{\textup C}$
is satisfied, where $T_{\textup C}$ is, with a very crude approximation, estimated as $T_{\textup C} \sim 1700$, and $M_{\textup C1}$ will be shown to be within zone $A'$ in this case with $T=600$.
Care that the above discussion does not mean that the Y-shaped trail is disadvantageous 
at $M \le M_{\textup C1}$, but means the long-term foraging efficiency $E_{\textup av}$ has same value 
between the V-shaped trail and the Y-shaped trail at $M \le M_{\textup C1}$. 
Therefore the value of $M$ at the zone boundary between $A$ and $A'$ (i.e. the boundary for the emergence of the V-shaped trail and the Y-shaped trail) reasonably falls below $M_{\textup C1}$.
Here it is remarked again that the transition 
from the V-shape to the Y-shape is geometrically continuous, however what is intrinsic is the emergence of a junction 
definitely separated from the nest by which the long-term efficient foraging is realized.
Furthermore it should be noted that the straightforward extension of the present model to the one with 
a larger number of feeding sites leads to a more striking transition from a superposition of simple straight trails
to the multi-branched trail as shown in fig. 6.  Such complicated branched patterns are seen in reality 
and are supposed to be related to the efficient foraging in a similar way to the present case.

With the larger amount of feeding at zone $B$, ants fails to make a {\it perfect foraging} with 
any shape of trails, while the Y-shaped trails like in zone $A'$ are kept formed. 
In this zone, it is neither easy to specify what is the optimized trail geometry,
nor is it easy to know the optimized foraging efficiency accompanied with the optimized trail.
Therefore, leaving a detailed qualitative analysis as the subject of further research 
we compare the obtained foraging efficiency by the present(un-constrainted) simulation in this zone $C$ as well as those in zones $A$ and $A'$, with that obtained by  
the geometrically constrained simulations in which the trail geometries are forced to be V. 
The latter was carried out through a specially arranged simulation with 
an appropriately tuned set of parameters on the pheromone evaporation rate and 
the pheromone secretion rate as written in tabel1. Notice that it is not an equivalent setup to the original (non-constrained)
simulation because of the tuning of the parameters, however we checked the above tuning 
causes, in the relevant range of $M$ ({\textup i.e.,} in zones $A,A'$ and $B$), 
only the deviation of trail geometry and that of the accompanying foraging efficiency 
from the original simulations. 
The last statement was verified by additional simulations under unsteady supply of food at a $\it single$ feeding site, 
in which the foraging efficiency obtained with the tuned sets of parameters was confirmed to be almost the same as that with the original set of parameters. 
Hence the averaged foraging efficiency for the constrained setup with the V-shaped trail was investigated 
in comparison with that of the original simulation, then we obtained the
outcome as shown in fig. 7.

In zone $A$ the averaged foraging efficiency is 
almost the same between the original simulation and the constrained one where 
in both simulations the {\it perfect foraging} is attained.
The same is the case up to $M=M_{\textup C1}$ in zone $A'$,
while over $M=M_{\textup C1}$ and up to the most right part zone $B$ the averaged foraging efficiency of the Y-shaped trail definitely exceeds that with the V-shaped trails, and at the most right part of zone $B$ the deviations again are 
very small.

In zone $C$, the geometrical symmetry of the trails breaks.
At the left edge of zone $C$ an intermediate shape 'y' between Y and / appears
which seems to be the outcome of the intermediate strategy between the last two,
and as the amount of supplied food increases, the pure /-shape is obtained.
The averaged foraging efficiency at the most right part of zone $C$ is $M/2T$, in other words, the slope of the $(\frac{M}{T},E_{\textup av})$ graph is 1/2 which indicates 
the {\it dedicated perfect foraging} meaning that ants are dedicated to carrying food of single feeding site.  
Like in zone $B$, the qualitative evaluation on the optimized foraging is hard to be performed. 
To make the matter worse, even the constrained simulation performed in zone $B$ 
will encounter several technical difficulties in this zone $C$.  
At least, in the most right part of zone $C$ where $T'+\tau >2T$ is satisfied, 
the expected formation time of the new trail leading to the alternative feeding site exceeds the 
absence time of food at the underlying feeding site, thus sticking to 
one feeding site without building an alternative trail is  reasonable for the 
efficient foraging. The further quantitative evaluation of the simulation from the viewpoint of the efficient foraging is yet to be made for zone $C$.

However, if the dynamics of ants is assumed to be within a strongly restricted frame,
the emergence of the /-shaped trail in zone $C$
is qualitatively interpreted as realizing an efficient foraging as well as the appearance of respective 
shapes of trail in zones $A, A', B$ and $D$.
Along this line, in the next section with a very simple approximation on the dynamics of the system,
we estimate the most efficient trail geometry among the V-shape, the Y-shape (here restricting to a 
symmetrical Y consisting of 
three equivalent lines attaching, respectively, to two feeding sites and a nest) and the /-shape 
according to the variation of the amount of supplied food. 

\section{Analysis}
In this section, using a simple approximation, we theoretically estimate the foraging efficiency $E_{\textup av}$ 
attained by the potentially most efficient foraging behavior as the function of $M/T$. The outcome will be compared with the results of simulations as explained above.
Here, the {\it basic period} 
of the foraging process with the V-shaped (or the Y-shaped) trail 
is defined as initiating at a feeding time at one feeding site 
and ending at the next feeding time at the alternative feeding site.
To make a direct comparison with the above simulation shown in fig. 5, the interval $T$ of this {\it basic period} is fixed as $T=600$. As the first main assumption of 
the present approximation, we consider each {\it basic period} is definitely divided into two 
parts, the first is i)the formation period of new trails during which ants are exploring new food sources or (re)building the new trail. In this period
food is not carried away from the feeding sites. The left part of the  {\it basic period} is ii)the exhausting period of food within which food is carried away with the potentially maximum foraging efficiency, 
$E_{\textup max}=a$ along the V-shaped trail, or $E_{\textup max}(Y)=a_Y$ along the Y-shaped trail.  
Then the temporal residual amount of food $M(t)$ at a feeding site in a
{\it basic period} varies so simply as shown in fig. 8(a-i).  Particularly at the end of each {\it basic period}, the residual amount of food at the recently feeding site is estimated as  
\begin{eqnarray}
M_{\textup res}^V(T) &=& 0 \qquad (\frac{M}{T}<(1-\frac{\tau}{T})a) \nonumber \\
             &=& M-(T-\tau)a  \qquad ((1-\frac{\tau}{T})a \le \frac{M}{T})  
\label{rv}
\end{eqnarray}
for the V-Shaped trail, while for the Y-shaped trail the corresponding quantity is estimated as,
\begin{eqnarray}
M_{\textup res}^Y(T) &=& 0 \qquad (\frac{M}{T}<(1-\frac{\tau_{\textup Y}}{T})a_{\textup Y}) \nonumber \\
             &=& M-(T-\tau_{\textup Y})a_{\textup Y} \qquad ((1-\frac{\tau_{\textup Y}}{T})a_Y \le \frac{M}{T})   
\label{rY}
\end{eqnarray}
as shown in fig. 8(a-ii).

As the second assumption of the present approximation we consider the same process is repeated at alternate feeding sites in every {\it basic period}, then the averaged foraging efficiency for the V-shaped trail can 
be expressed as 
$E_{\textup av}(V)= (M-M_{\textup res}^V(T))/T$.
So that,
\begin{eqnarray}
E_{\textup av}(V) &=& M/T  \qquad  (\frac{M}{T}<(1-\frac{\tau}{T})a) \nonumber \\
          &=& (1-\tau/T)a = (1-\tau/T)E_{\textup max} \qquad ((1-\frac{\tau}{T})a \le \frac{M}{T}),
\label{ev}
\end{eqnarray}
while for the Y-shaped trails,
\begin{eqnarray}
E_{\textup av}(Y) &=& M/T   \qquad (\frac{M}{T}<(1-\frac{\tau_{\textup Y}}{T})a_{\textup Y}) \nonumber \\
          &=& (1-\tau_{\textup Y}/T)a_{\textup Y} = (1-\tau_{\textup Y}/T)E_{\textup max}(Y) \qquad ((1-\frac{\tau_{\textup Y}}{T})a_{\textup Y} \le \frac{M}{T})
\label{eY}
\end{eqnarray}
is satisfied. These results are shown in fig. 9 with the following estimation of specific values 
of relevant quantities.
Comparing eq.(\ref{ev}) with eq.(\ref{eY}), the Y-shaped trail is seen more advantageous than the V-shaped trail
at $M> (T-\tau)a$ as long as the relation  
\begin{equation}
T<T_C \equiv \frac{\tau a-\tau_Y a_Y}{a-a_Y}
\label{tc}
\end{equation}
holds, where $T_{\textup C}$ will be estimated in the below as $T_{\textup C} \sim 1700$ over which the emergence of the Y-shaped trail 
causes no benefit to ants. In such cases, the direct transition from the V-shape to the /-shape at a certain $M$ is expected to appear for the efficient foraging. This will be discussed later.

Moreover, for the cases with the /-shaped trail, one of the two feeding sites is completely ignored by ants.
Thus the {\it extended basic period} with interval $2T$ is defined in place of the {\it basic period},
which initiates at a feeding time of 
a feeding site and ending at the 2nd next feeding taking places at the same feeding site, and this period also is assumed to be divided into i) the trail formation period and ii) the exhausting period.
Here $\tau$, the trail formation time, is assumed to be the same as that of the V-shaped trail considering
the straightness of this trail. In the same reason the potentially maximum foraging efficiency using the /-shaped trail is assumed like $E_{\textup max}(/)=E_{\textup max}=a$.
Furthermore, the whole dynamics of the system is assumed to consist of the
simple repetition of the equivalent {\it extended basic periods}(fig. 8(b-i)).
Care this assumption is found too much artificial in a certain case thus will be refined 
depending on the case.
Below, with the above approximation, we estimate the foraging efficiency attained with the /-shaped trail. 
Firstly, if $M_{\textup res}^/(2T)$=0, namely, the residual amount of 
food at the end of the {\it extended basic period} is zero, 
the averaged foraging efficiency satisfies $E_{\textup av}(/)=M/2T$ because, in this case, ants completely ignore the alternative feeding site while making the {\it dedicated perfect foraging} insisting on 
the underlying feeding site.    
This occurs when $M < (2T-\tau)a$.
On the other hand, if $M_{\textup res}^/(2T) >0$, that is, if excessive amounts of food over $M=(2T-\tau)a$ are supplied, they are not completely carried away within an {\it extended basic period}, however if the dynamics of the system obeys the above assumption, the next {\it extended basic period} will 
initiate again from the trail formation period in spite of the established trail, 
and simple repetition of the previous {\it extended basic period} takes place.
Then the averaged foraging efficiency will no more increase over  $M/2T=(1-\tau/2T)a$.
This is an unnatural situation.  Therefore we make a simple refinement of the present approximation so that 
the trail formation period is omitted in an {\it extended basic period} if  
food remains unexhausted at the end of the previous {\it extended basic period} (fig. 8(b-ii)).
That means, at $(2T-\tau)a \le M < (2Ta)$, we consider 
{\it a pair of extended basic periods} as the unit interval of the system,
which consists of the former half containing both 
the trail formation period and the exhausting period and the latter half
free of the trail formation period(fig. 8(b-ii))).
Note that at end of the former half, a finite amount of food remains unexhausted at the feeding site, while before the end of the latter all the food is carried away. 
With more food such that $2Ta \le M$, at the end of every {\it extended basic period},
food always remains at the feeding site to which the trail is steadily connected, so ants can 
spend all of the time for carrying food without wasting time for the
trail formation as shown in fig. 8(b-iii). Therefore $E_{\textup av}(/)=E_{\textup max}$ is attained in this case.   
The foraging efficiency obtained by the /-shaped trail is summarized in a short expression,
\begin{eqnarray}
E_{\textup av}(/) &=& M/2T \qquad (\frac{M}{T} < (2-\frac{\tau}{T})a) \nonumber\\
       &=& M/4T+(2T-\tau)a/4T  \qquad ((2-\frac{\tau}{T})a \le \frac{M}{T} < 2a) \nonumber\\
      &=& a=E_{\textup max}  \qquad ( 2a \le \frac{M}{T}).
\end{eqnarray}
The above result is shown in fig. 9 with the following quantitative estimation of relevant quantities.

Now, the specific values of $\gamma$,  $\tau$ , $\tau_Y$, $a$ ,$a_Y$,$M_{\textup C1}$
and $M_{\textup C2}$ in the present model are roughly estimated through a combination 
of the above discussions and the outcome of simulations. 
Firstly, $M_{\textup C1}$, the critical amount of $M$ over which the V-shaped trail
fails to make a {\it perfect foraging}, is described through the relation eq.(\ref{ev}) as $M_{\textup C1}=(T-\tau)a$.  
In fig. 7 it corresponds to the point where the constrained and the unconstrained simulations 
begin to exhibit a different $E_{\textup av}$, that is, $M_{\textup C1} \sim 1000$.  On the other hand $a$ is directly estimated from the relation $a=E_{\textup max}$ which corresponds to the saturated foraging 
efficiency, $E_{\textup av}\sim 6$, in zone $D$ in fig. 5. The last
estimation directly leads to $\gamma \sim 0.72$ using the relation
$a=\gamma \frac{Nmv}{2L}$. Hence considering $T=600$ for fig. 5 and 
eq.(\ref{ev}), $\tau \sim 430$ is obtained. 
In a similar way $M_{\textup C2}$, the critical amount of $M$ at which the Y-shaped trail
fails to make a {\it perfect foraging}, is expressed through
eq.(\ref{eY}) as $M_{\textup C2}=(T-\tau_{\textup Y})a_{\textup Y}$.
This value is estimated from fig. 5 as $M_{\textup C2} \sim 2000$ and with the relations  
$a_{\textup Y} \equiv\gamma\frac{Nmv}{2L'}$,
$a \equiv \gamma\frac{Nmv}{2L}$ and $L=\cos 30^{\circ}L'$,  relations $a_{\textup Y} \sim 5.1 $ and $\tau_{\textup Y} \sim 210$ are obtained.
Subsequently $T_{\textup C}$ in eq.(\ref{tc}) is estimated as $T_{\textup C} \sim 1700$.

On the basis of the above discussions, we estimated the the relations for $(\frac{M}{T},E_{\textup av})$ at $T=600$
for the respective types of the trails, and depicted in fig. 9.
In this figure, if we trace the the most efficient trail geometry 
with the increase of $M$, the corresponding trail varies from the V-shaped trail(equivalently Y-shaped trail) 
to the Y-shaped trail and finally to the /-shaped trail which
qualitatively follows the variation of the emergent trail geometries in
the simulation as shown in fig. 5. 
Remark that according to eq.(\ref{tc}), at $T>T_{\textup C}$  Y-shaped trail has no advantage in any 
value of $M$, hence a direct transition from 
the V-shaped trail to the /-shaped trail is expected.  
Actually in fig. 4 at the columns with $T>1700$ no definite Y-shaped trail is seen though there are imperfect Y-shapes with very short vertical parts, but they are supposed to deform again into the /-shaped trail with the increase of $M$. Further calculations are required to confirm this conjecture.
Remember these analytical results are based on a crude approximation ignoring the complicated 
dynamics exhibited in the simulation like the aperiodic time evolution of trails which probably 
is the origin of the more efficient foraging than the theoretically 
expected one. More detailed discussions are required to allow an quantitative understanding of the system.

\section{Summary}
Using a simple model of ants we showed that they, as a group, 
perform a flexible foraging by coping with the variation 
of the feeding schedule through the alternation of the trail geometry.
Especially in the regime of intermediate amount of food, less than sufficient amount necessary to offer a permanent food source but more than barely enough for the temporal emergence of straight trail,
ants build branched trails (the Y-shaped trails) which realize a long-term efficient foraging in spite of 
the short-term redundancy. 
In addition, holding the Y-shaped trail regime in between,
the V-shaped trails and the /-shaped trails leads 
to make the efficient foraging in the respective situations.
Notice that if the number of food sources is increased, a more spectacular transition of trail geometry 
will be seen as shown in fig. 5  
which corresponds to more realistic situations accompanied by 
more complicated foraging strategies, though the essential reasoning from the present study would remain unchanged.
We have yet to investigate further aspects like how such an efficient strategy is attained only by the local dynamics of 
individual ants with a small set of relevant parameters and to what extent this mechanism is robust under 
the variation of the underlying parameters.
Of course, the comparison of 'our' ants to 
real ants is required though several tough problems should be overcome,
like the specification of 
pheromones relevant to the trail formation,
also linking the present reasoning to the group dynamics of 
robots could be one of the potential applications \cite{Bo}.

\bigskip
\noindent

The authors thank the participants of the research project 
``Creation and Sustenance of Diversity'' organized by the International Institute of Advanced Studies, Kyoto.
We also appreciate Hans-Georg Mattutis for critical reading of the manuscript.

\newpage

{Figure Captions}

\vspace{.2in}

\noindent
FIG 1  Each ant is, at each time step, 
heading to one of the six nearest sites.
This heading direction corresponds to the moving direction in the previous
time step. The possible directions in each time step are:
the forward (=facing) direction and its neighboring (right and left)
directions. 

\medskip
\noindent
FIG 2 The method of feeding. 
Feeding sites are located at two corners of an
equilateral triangle, the 3rd corner of which is the nest site.
The amount $M$ of food is supplied, 
in turn, from alternative feeding sites at every feeding interval $T$. 
%After each feeding event, they decreases as taken away by ants.

\medskip
\noindent
FIG 3 Typical shapes of trails obtained in the simulation.
i)V-shaped trail, ii)Y-shaped trail, iii)/-shaped trail.
Here dark shading in each figure 
means the density of the ants averaged over a time significantly longer than $T$, the interval between 
successive feeding events.

\medskip
\noindent
FIG 4 The relation between emergent trail patterns 
and the combination of feeding schedule parameters,
\{$M$,$T$\}. The dark shading in each figure 
means the density of ants averaged over a time significantly longer 
than $T$ and over the ensemble of 5 simulations for the V-shaped and the Y-shaped trails.
For the /-shaped trail the ensemble average is not taken because of their intrinsic asymmetry.

\medskip
\noindent
FIG 5 The relation between the 
supplied amount of food per unit time $M/T$ and the long-term averaged foraging efficiency $E_{av}$.
The figure at the bottom is the inset of the above.
There are five characteristic zones $A, A', B, C$ and $D$.
In zone $A$, the V-shaped trail appears, while in zones $A',B$ 
the Y-shaped trail appears, and in zones $C$ and $D$ the /-shaped trail is 
seen. Up to the zone $A'$ the {\it perfect foraging} is 
attained whereas in zone $D$ the amount of supplied food exceeds the 
carrying capacity of ants thus $E_{av}$ reaches a saturation value $a$.

\medskip
\noindent
FIG 6 Typical trail pattern 
in the case with four feeding sites.

\medskip
\noindent
FIG 7 The relation for $(\frac{M}{T},E_{av})$ explained in fig.5 as indicated by $+$,
while $\times$ shows the same relation obtained through a constrained simulation 
in which only the V-shaped trail can be built.
Over $M=M_{C1}$ the Y-shaped trail strategy becomes more advantageous than 
the V-shaped trail strategy, and this situation prevails until the right edge of the zone $B$.

\medskip
\noindent
FIG 8 a)(i)The simplified time evolution of $M(t)$, the residual amount of food 
at feeding sites. 
At $t=0$, the amount $M$ of food is supplied at one feeding site.
In the beginning, they are not exhausted because ants are at the process of 
exploring food and/or building a trail, this period is denoted as $\tau$ for the V-shaped trail (or $\tau_Y$ for the Y-shaped trail), and only after a trail to the feeding site is established, food can be carried away with the decreasing rate $a$ for the V-shaped trail
(or $a_Y$ for the Y-shaped trail). 
In the present approximation, after every feeding time $t=nT, \{n=0,1,2, \dots \}$
the equivalent process is repeated at alternative feeding sites, so 
the span between $t=nT$ and $t=(n+1)T$ is called the {\it basic period}.
(ii)$M_{res}(T)$ is defined as the residual amount of food at the end af a {\it basic period}.
This quantity deviates between that for the V-shaped trail and that for the Y-shaped trail.

\noindent
b)(i)For the foraging with the /-shaped trail, the length of the {\it basic period} of the system is extended to $2T$ which begins at $t=2nT$ and ends at $t=2(n+1)T$ because one feeding site 
is completely ignored in this regime. We call this time span {\it the extended basic period}.  

\noindent
b)(ii)However, if the residual food at $2nT$ is not zero, the 
currently working trail is continuously used to save the new trail formation time $\tau$ after the 
supply of new food at the same instance. Then the $basic period$ is extended to $4T$.  

\noindent
b)(iii) With a larger amount of food supply than the above, the $basic period$ comes again to $2T$.

\medskip
\noindent
FIG 9 The theoretically estimated relation between $M/T$ and averaged foraging 
efficiency $E_{av}$ for the respective trail shapes: $E_{av}(V)$ for the V-trail with 
the real line, $E_{av}(Y)$ with the dotted line,
and $E_{av}(/)$ with the broken line.
According to these relations, the most efficient shape of trail varies, 
as $M/T$ increases, from 
V(or Y) to Y, and finally the /-shape is seen most efficient. This trend corresponds to the emergent trail 
shapes in our simulation with T=600.
Note this is the output of a crude approximation, so it contains qualitative deviations
compared to those obtained in the simulation shown in fig.5.

\vspace{15pt}

\noindent
{Table Captions}

\medskip
\noindent Table 1  Specific values of parameters used in simulations.


\begin{thebibliography}{99}
\bibitem{trail}
It should be remarked that, the trunk trail keeps its shape even in the period where food is absent 
as long as $\tau$ (the trail formation time to new food) is shorter than the residual time of the 
foot pheromone which is set to be satisfied in the present simulation. The detailed discussion taking these time scales into account will be the subject of the following publication.
\bibitem{Diff} 
Note this diffusion process is performed only at the occupied sites by ants at the corresponding time step.
This artificial diffusion is introduced to smooth out the pheromone density field.
In addition, through the diffusion the non-local (but not the long-range) 
perception of the evaporated pheromones by ants is 
effectively realized. The refinement of this rule into more sophisticated form
wil be left for further investigation.
\end{thebibliography}
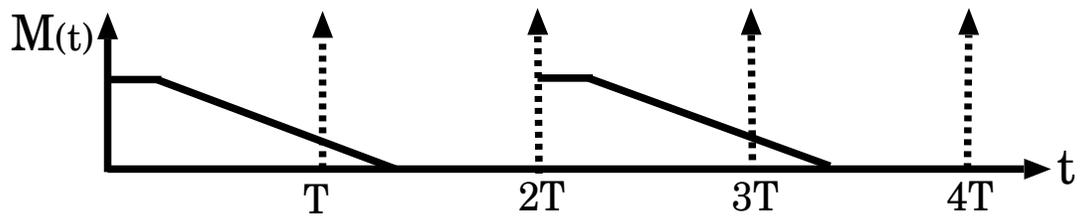
\end{document}